\documentclass[lettersize,journal]{IEEEtran}
\usepackage{amsmath,amsfonts}
\usepackage{amssymb}
\usepackage{algorithm,algpseudocode}
\usepackage{array}
\usepackage[caption=false,font=normalsize,labelfont=sf,textfont=sf]{subfig}
\usepackage{textcomp}
\usepackage{stfloats}
\usepackage{url}
\usepackage{verbatim}
\usepackage{graphicx}
\usepackage{cite}
\usepackage{color}
\def\BibTeX{{\rm B\kern-.05em{\sc i\kern-.025em b}\kern-.08em
		T\kern-.1667em\lower.7ex\hbox{E}\kern-.125emX}}
\usepackage{balance}
\usepackage{epstopdf}
\begin{document}
	
	\title{Generative Learning Powered Probing Beam Optimization for Cell-Free Hybrid Beamforming}
	
	\author{Cheng Zhang,~\IEEEmembership{Member,~IEEE,}
		Shuangbo Xiong, Mengqing He, Lan Wei,\\
		Yongming Huang,~\IEEEmembership{Senior Member,~IEEE}, Wei Zhang,~\IEEEmembership{Fellow,~IEEE} \vspace{-1cm}
		\thanks{
			This work was supported in part by the National Natural Science Foundation of China under Grant 62271140, 62225107,  the Natural Science Foundation of Jiangsu under Grant BK20240174, BK20222001, the Fundamental Research Funds for the Central Universities 2242022k60002, and the Major Key Project of PCL. $\textit{(Corresponding authors: Cheng Zhang, Yongming Huang.)}$
			
			C. Zhang, S. Xiong, M. He, L. Wei, and Y. Huang are with the National Mobile Communication Research Laboratory, Southeast University, Nanjing 210096, China, and also with the Purple Mountain Laboratories, Nanjing 211111, China (e-mail: zhangcheng\_seu@seu.edu.cn, shuangbo\_xiong@seu.edu.cn, hemengqing1997@163.com, weilan@seu.edu.cn, huangym@seu.edu.cn).
			
			W. Zhang is with the School of Electrical Engineering and Telecommunications,
			the University of New South Wales, Sydney, NSW 2052, Australia, and
			also with the Purple Mountain Laboratories, Nanjing 211111, China (e-mail:
			wzhang@ee.unsw.edu.au).}
	}

	
	\maketitle
	
	\begin{abstract}
		Probing beam measurement (PBM)-based hybrid beamforming provides a feasible solution for cell-free MIMO. In this letter, we propose a novel probing beam optimization framework where three collaborative modules respectively realize PBM augmentation, sum-rate prediction and probing beam optimization. Specifically, the PBM augmentation model integrates the conditional variational auto-encoder (CVAE) and mixture density networks and adopts correlated PBM distribution with full-covariance, for which a Cholesky-decomposition based training is introduced to address the issues of covariance legality and numerical stability. Simulations verify the better performance of the proposed augmentation model compared to the traditional CVAE and the efficiency of proposed optimization framework.
	\end{abstract}
	
	\vspace{-0.8cm}
	
	\section{Introduction}
	\IEEEPARstart{C}{ell-Free} (CF) MIMO systems provide significant performance improvement for wireless interface through the collaboration among multiple access points (APs) assisted by the central unit (CU) \cite{wang2023full,10183795}. 
	However, with the increase in the number of collaborative APs, the system overhead for the acquisition of channel state information (CSI) and the complexity for cooperative beamforming (BF) both become unaffordable \cite{feng2021weighted}, especially for fast time-varying environments, e.g., vehicular networks.
	
	Hybrid BF has shown its potential in achieving near-optimal performance via using only a small number of radio frequency (RF) chains for CF MIMO systems \cite{wang2022joint}. 
	Recently, deep learning (DL) technologies have been utilized to reduce the computational complexity \cite{yetis2021joint} and CSI overhead \cite{hojatian2022decentralized} for CF MIMO hybrid BF, respectively.  
	The probing beam measurement (PBM) is the partial beam-domain representation of complete CSI, with relatively small acquisition overhead \cite{ma2020machine, alkhateeb2018deep}.  
	For the PBM-based BF design, the optimization of probing beams is of significant importance. In \cite{heng2022learning}, an end-to-end deep neural network (DNN) was designed to jointly learn the probing codebook and the beam predictor for single-cell analog BF, where the user distribution or channel realization is used as the prior condition. In practical scenarios, this prior information does not necessarily exist or is not accurate. In CF MIMO systems, the number of probing beam combinations grows exponentially with the number of APs. These both result in insufficient data samples for probing beam optimization.
	
	Existing data augmentation models for PBM can be divided into two categories.
	One is to use DNN to establish a deterministic mapping between fingerprint information, e.g., location, and PBM, so as to predict the PBM of unsampled interesting locations via interpolation or extrapolation \cite{sun2018augmentation}, \cite{he2021intelligent}. 
	Another is to use generative models, e.g., generative adversarial networks (GAN) \cite{njima2021indoor} and conditional variational auto-encoder (CVAE) \cite{chen2020progressive} to learn the distribution of PBM based on limited PBM samples and then use it to realize PBM augmentation, where the major application scenario is the indoor positioning \cite{njima2021indoor, chen2020progressive}. The GAN \cite{ye2020deep} and VAE \cite{li2021influence} models have also been adopted to solve the problem of CSI insufficiency in the training of DNN models for decoding and power control in wireless communications, respectively.
	
	In this letter, we study the optimization of probing beam for PBM-based hybrid BF in CF MIMO systems. Specifically, the PBM-based hybrid BF used is the one we proposed in our previous work \cite{10255707}, where the DNN exploits the PBM to compress the joint analog beamspace with unsupervised learning, based on which the analog BF is conducted efficiently with typical beam selection algorithms, e.g., the strongest beam first (SBF) \cite{orikumhi2020sinr}, and the linear minimum mean square error (LMMSE)-type digtal BF is then obtained based on beam-domain CSI. Main contributions are given as follows.
	\begin{itemize}
		\item
		A probing beam optimization framework is proposed where a PBM augmentation module and a DNN-based sum-rate predictor, respectively, create sufficient PBMs and predicted sum-rate of the hybrid BF scheme, based on which standard optimization algorithms, e.g., the genetic algorithm \cite{mirjalili2019genetic}, optimizes the probing beam configuration.
		\item
		The proposed PBM augmentation model organically integrates the CVAE and mixture density networks (MDN) to improve the augmentation accuracy. And we use the MDN with full covariance matrix rather than the diagonal one to exploit the PBM correlation between different probing beams and users, for which a Cholesky-decomposition based training method is proposed to address the legality issue of the covariance matrix and the numerical stability problem in high dimensions.
		\item 
		Simulation results show that the proposed PBM augmentation model learns more accurate PBM distribution compared to the traditional CVAE augmentation model and is able to create PBMs for unsampled probing beam combinations. In addition, the proposed probing beam optimization framework can realize efficient probing beam optimization with limited PBM samples. 
	\end{itemize}

	\vspace{-0.3cm}
	\section{System Model}
	\vspace{-0.1cm}
	A CF MIMO system is considered where $B$ APs each with $M$ antennas and $U$ RF chains which are connected to the CU via fronthual links serve $U$ single-antenna users cooperatively.
	
	\vspace{-0.4cm}
	\subsection{Channel Model}\label{Channel Model}
	\vspace{-0.1cm}
	The downlink channel $\boldsymbol{h}_{b, u}\in \mathbb{C}^{M\times1}$ from the AP $b\in \mathbb{B}=\{1,... ,B\}$ to the user $u\in \mathbb{U}=\{1,... ,U\}$ is \cite{alkhateeb2019deepmimo}
	\setlength\abovedisplayskip{1pt}\setlength\belowdisplayskip{1pt}
	\begin{equation}
		\hspace{-0.05cm}\boldsymbol{h}_{b, u} = \sum_{l=1}^{L} \hspace{-0.05cm}\sqrt{\rho_{l}^{b, u}} e^{j 2 \pi \tau_{l}^{b, u} W} \mathbf{a}\hspace{-0.05cm}\left(\phi_{\mathrm{az},l}^{b, u}, \phi_{\mathrm{el},l}^{b, u}\right), \label{channel} 
	\end{equation}
	where $L$ is the path number. $\rho_{l}^{b, u}$, $ \tau_{l}^{b, u}$, $W$, $\phi_{\mathrm{az},l}^{b, u}$ and $\phi_{\mathrm{el},l}^{b, u}$ are the path gain, the delay, the bandwidth, the azimuthal angle and the elevation angle of the $l$-th path, respectively. $\mathbf{a}(\cdot)$ is the array response vector.
	For a rectangular uniform array with half-wavelength antenna spacing, 
	$\mathbf{a}\left(\phi_{\mathrm{az},l}^{b, u}, \phi_{\mathrm{el},l}^{b, u}\right)= \mathbf{a}_\mathrm{y}\left(\phi_{\mathrm{az},l}^{b, u}, \phi_{\mathrm{el},l}^{b, u}\right) \otimes \mathbf{a}_\mathrm{z}\left(\phi_{\mathrm{el},l}^{b, u}\right)$
	with $\mathbf{a}_\mathrm{y}\left(\phi_{\mathrm{az},l}^{b, u}, \phi_{\mathrm{el},l}^{b, u}\right)=\left[1,\ldots, e^{j \pi\left(M_\mathrm{y}-1\right) \sin \left(\phi_{\mathrm{el},l}^{b, u}\right) \sin \left(\phi_{\mathrm{az},l}^{b, u}\right)}\right]^\mathrm{T}$ 
	and $\mathbf{a}_\mathrm{z}\left(\phi_{\mathrm{az},l}^{b, u}, \phi_{\mathrm{el},l}^{b, u}\right)=\left[1,\ldots, e^{j \pi\left(M_\mathrm{z}-1\right) \cos \left(\phi_{\mathrm{el},l}^{b, u}\right)}\right]^\mathrm{T}$ and $M_\mathrm{y}$ and $M_\mathrm{z}$ ($M = M_\mathrm{y}\times M_\mathrm{z} $) denoting the number of antennas at the horizontal direction and vertical direction, respectively.
	
	\vspace{-0.45cm}
	\subsection{Transmission Model}
	\vspace{-0.1cm}
	Consider the hybrid BF design with the analog BF matrix of AP $ b\in\mathbb{B}$, i.e., $\boldsymbol{F}_{{\rm RF}, b}=\left[\boldsymbol{f}_{i_{b, 1}} \ldots \boldsymbol{f}_{i_{b, U}}\right]\in \mathbb{C}^{M\times U}$, and the digital BF vector $\boldsymbol{w}_{\mathrm{BB},u}\in \mathbb{C}^{BU\times 1}$ for user $u\in\mathbb{U}$, 
	where the analog beam $\boldsymbol{f}_{m}\in \mathbb{C}^{M\times 1}$, $m\in \mathbb{I}_{b}=\left\{i_{b, 1}, \ldots, i_{b, U}\right\}, \forall b\in\mathbb{B}$ is chosen from the analog beam codebook $\boldsymbol{F}\in \mathbb{C}^{M\times M }$, e.g., the standard discrete Fourier transform (DFT) codebook. 
	Define $\boldsymbol{h}_u = \left[\boldsymbol{h}_{1,u}^{\mathrm T},...,\boldsymbol{h}_{B,u}^{\mathrm T}\right]^\mathrm{T}\in \mathbb{C}^{BM\times 1}$, $\boldsymbol{H} = \left[\boldsymbol{h}_1,...,\boldsymbol{h}_U\right]\in \mathbb{C}^{BM\times U}$, the received signal of the user $u$ is
	\setlength\abovedisplayskip{1pt}\setlength\belowdisplayskip{1pt}
	\begin{equation} 
		y_u = \boldsymbol{h}_u^\mathrm{H} \boldsymbol{F}_\mathrm{RF}\boldsymbol{w}_{\mathrm{BB},u}s_u + \sum_{v\neq u}\boldsymbol{h}_u^\mathrm{H} \boldsymbol{F}_\mathrm{RF}\boldsymbol{w}_{\mathrm{BB},v}s_v + n_u,
	\end{equation}
	where $s_u$ is the data symbol for user $u$ with unit power, and $n_u \sim \mathcal{C} \mathcal{N}(0,\sigma_{u}^{2})$ denotes the receiver additive Gaussian noise. $\boldsymbol{F}_\mathrm{RF}=\mathrm{diag}\left\{\boldsymbol{F}_{\mathrm{RF},1},...,\boldsymbol{F}_{\mathrm{RF},B}\right\}\in \mathbb{C}^{BM\times BU}$. The achievable rate of user $u$ can be written as
	\setlength\abovedisplayskip{1pt}\setlength\belowdisplayskip{1pt}
	\begin{equation}\label{rate}
		R_u = \log_2\left(1+\frac{\vert\vert\boldsymbol{h}_u^\mathrm{H}\boldsymbol{F}_{\rm RF}\boldsymbol{w}_{\mathrm{BB},u}\vert\vert_2^2}{\sum_{v\ne u}\vert\vert\boldsymbol{h}_u^\mathrm{H}\boldsymbol{F}_{\rm RF}\boldsymbol{w}_{\mathrm{BB},v}\vert\vert_2^2+\sigma_u^2}\right).
	\end{equation}

	\vspace{-0.4cm}
	\subsection{Probing Beam Powered Hybrid BF}\label{PBHBF}

	If the analog beams of AP $b$, i.e., $\boldsymbol{f}_{m}$, $m\in \mathbb{I}_{b}$, are directly chosen from the standard DFT codebook $\boldsymbol{F}$, the analog beam training overhead is unaffordable due to the large number of APs and AP antennas in our considered CF MIMO systems. Recall that the PBMs from multiple APs provide a rough description of the user location, on which the beam-domain CSI is highly dependent \cite{alkhateeb2018deep}. Therefore, the PBMs from multiple APs are utilized to compress the beamspace for analog beam training as in our previous work \cite{10255707}.  
	
	Specifically, each AP $b\in\mathbb{B}$ first chooses $N_b$ probing beams with beam indices of $\tilde{\mathbb{I}}_{b}=\left\{\tilde{i}_{b, 1}, \ldots, \tilde{i}_{b, N_b}\right\}$ from the probing beam codebook $\tilde{\boldsymbol{F}}_b$ and obtains the PBMs $\boldsymbol{r}_{b,u},u\in\mathbb{U}$ with ${r}_{b,u,i} = \vert\tilde{\boldsymbol{f}}_{b,i}^\mathrm{H}\boldsymbol{h}_{b,u}\vert^2, \forall i\in\tilde{\mathbb{I}}_{b}$. Then, the CU collects these PBMs from each AP and uses a DNN to obtain the compressed analog beam codebook for each AP $b\in\mathbb{B}$, which consists of analog beams $\boldsymbol{f}_{m}$, $m\in \mathbb{A}_b=\{a_{b,1},...,a_{b,|\mathbb{A}_b|}\}$. 
	Finally, each AP $b\in\mathbb{B}$ trains these analog beams $\boldsymbol{f}_{m}$, $m\in \mathbb{A}_b$ to obtain beam-domain CSI $\{\boldsymbol{h}_{b,u}^\mathrm{H}{\boldsymbol{f}}_{m},\forall u\in\mathbb{U}, m\in\mathbb{A}_b, b\in\mathbb{B}\}$, based on which  typical beam selection approaches, e.g., the strongest beam first (SBF) \cite{orikumhi2020sinr}, can be conducted to output the analog beams with indices $\mathbb{I}_{b}=\left\{i_{b, 1}, \ldots, i_{b, U}\right\}, \forall b\in\mathbb{B}$ for the hybrid BF. 
	For the digital part $\boldsymbol{W}_\mathrm{BB}=\left[\boldsymbol{w}_{\mathrm{BB},1},...,\boldsymbol{w}_{\mathrm{BB},U}\right]\in \mathbb{C}^{BU\times U}$ in the hybrid BF, the LMMSE one is 
	\setlength\abovedisplayskip{1pt}\setlength\belowdisplayskip{1pt}
	\begin{equation}
		\boldsymbol{W}_\mathrm{BB}=\bar{\boldsymbol{H}}\left(\bar{\boldsymbol{H}}^\mathrm{H} \bar{\boldsymbol{H}}+\lambda \mathbf{I}_{U}\right)^{-1} \boldsymbol{\Sigma},\label{WBB}
	\end{equation}
	where $\bar{\boldsymbol{H}}=\boldsymbol{F}_{\rm RF}^{\mathrm H}\boldsymbol{H}\in \mathbb{C}^{BU\times U}$. $\boldsymbol{\Sigma}=\mathrm{diag}\{\rho_1,...,\rho_U\}\in \mathbb{R}^{U\times U}$ with $\rho_u={\sqrt{P_u}}/{\vert\vert\bar{\boldsymbol{w}}_{\mathrm{BB},u}\vert\vert_2}$ where   $\bar{\boldsymbol{W}}_\mathrm{BB}=\boldsymbol{F}_{\mathrm{RF}}\bar{\boldsymbol{H}}\left(\bar{\boldsymbol{H}}^\mathrm{H} \bar{\boldsymbol{H}}+\lambda \mathbf{I}_{U}\right)^{-1}\in \mathbb{C}^{BM\times U}$ and $P_{u}$ is the power allocated for user $u\in\mathbb{U}$.  
	
	For the PBM based beamspace compression, the DNN model and its training mechanism are introduced in the following. 
	The DNN model input is the set of PBMs, i.e., $\{r_{b,u,i}, u\in \mathbb{U}, i\in{\tilde{\mathbb{I}}_b}, b\in\mathbb{B}\}$.
	To exploit the correlation among adjacent users' beam-domain channels and that among neighboring beams of the same user, the PBMs are organized into $B$ matrices $\boldsymbol{R}_b, \forall b\in\mathbb{B}$ with $[\boldsymbol{R}_b]_{u,i}=r_{b,u,i}$. The hidden layers are mainly composed of convolutional units which significantly reduces the number of network parameters by exploiting the channel correlation.
	To balance the learning difficulty and the beamspace suppression efficiency, we classify the complete beamspace of each AP into $C$ classes based on the spatial relationship of the beams, i.e., $\mathbb{A}_b\in\left\{\mathbb{A}_b^{(1)},...,\mathbb{A}_b^{(C)}\right\}$. Then, the softmax layer is adopted for the $B\times C$ dimensional output $\boldsymbol{y}$  and $\mathbb{A}_b = \mathbb{A}_b^{(i_b^\star)}, \forall b\in\mathbb{B}$ with $i_b^\star = \arg \max_{i\in\{1,...,C\}} y_{(b-1)\times C+i}$.
	Since the label of the DNN model for beamspace compression is difficult to obtain, the unsupervised training method is adopted as in \cite{10255707}, \cite{hojatian2021unsupervised}. Specifically, the loss function is defined as 
	\begin{equation}
		\mathrm{Loss}_{0} = -\alpha f(R_\mathrm{sum})\sum_{b=1}^{B}y_{(b-1)\times C+i_b^\star},
	\end{equation}
	where $R_\mathrm{sum}$ is calculated from Eq. \eqref{rate} based on the  beam selection in compressed beamspaces $\mathbb{A}_b, \forall b\in\mathbb{B}$. $f(R_\mathrm{sum})$, e.g.,  $R_\mathrm{sum}^2$, is a function that controls the sensitivity of the loss to $R_\mathrm{sum}$. $\alpha$ is the hyperparameter that controls the range of loss values. Standard optimization methods are then used to reduce the training loss. In the training stage, the complete beam-domain CSI $\{\boldsymbol{h}_{b,u}^\mathrm{H}{\boldsymbol{f}}_{i}, \forall u\in\mathbb{U}, i\in\mathbb{M}, b\in\mathbb{B}\}$ is needed for evaluating the performance of the DNN compression model, while in the execution stage, only PBMs are needed.
	

	\vspace{-0.5cm}
	\section{Generative Learning Powered Probing Beam Optimization}
	\vspace{-0.2cm}
	For the PBM-based beam selection, the optimization of probing beam is important. 
	Different from real-time beam selection based on beam training, the  probing-beam optimization is generally conducted in greater time granularity, e.g., a time interval during which the randomness of CSI should be considered. Therefore, the optimization problem can be formulated as a combinational optimization problem
	\begin{equation}\label{Eq15}
		{\max _{\tilde{\mathbb{I}}_{b}, \forall b \in \mathbb{B}}} {\mathbb{E}}_{\boldsymbol{H}}\left\{R_\mathrm{sum} \mid \tilde{\mathbb{I}}_{b}, \forall b \in \mathbb{B}\right\}.
	\end{equation}
	For better clarification, we assume that there are $L^\mathrm{total}$ probing beam combinations, denoted as $\tilde{\mathbb{I}}^{(l)} = \{\tilde{\mathbb{I}}_{1}^{(l)},...,\tilde{\mathbb{I}}_{b}^{(l)}\},l=1,...,L^\mathrm{total}$. Define $\mathbb{L}^\mathrm{total}=\{1,...,L^\mathrm{total}\}$.
	
\begin{figure}[htbp]
		\vspace{-0.4cm}
		\centering
		\setlength{\abovecaptionskip}{0.cm}
		\includegraphics[scale=0.32]{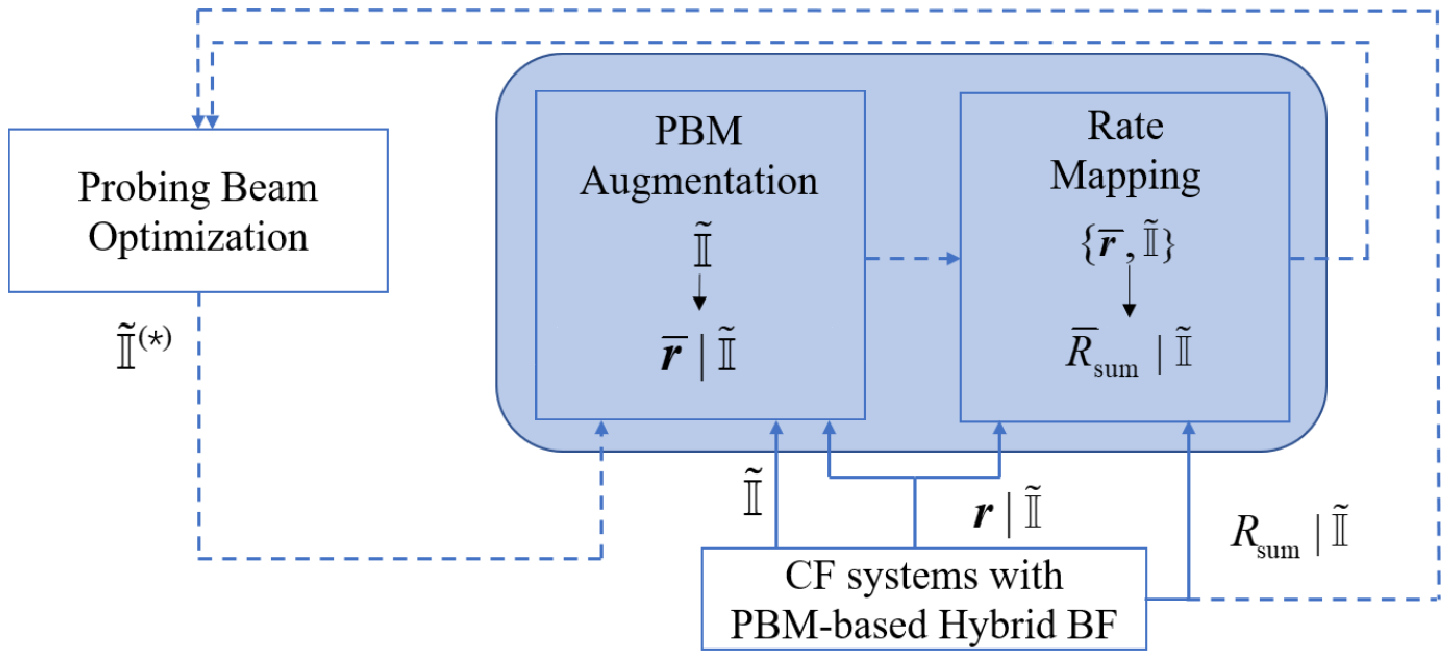}
		\caption{Framework for probing beam optimization. The solid and dotted lines correspond to the training and execution phase.}\label{ssb opt}
		\vspace{-0.4cm}
	\end{figure}

	In CF MIMO systems, the solution space for Problem Eq. \eqref{Eq15} has an exponential increase with the AP number $B$. Due to the cost of data acquisition, it is difficult for practical systems to obtain data samples for all possible probing beam combinations and multi-user channel realizations.
	In addition, arbitrary adjustment of probing beam configuration may bring severe fluctuations in network performance.
	Therefore, we propose an optimization framework as shown in Fig. \ref{ssb opt}.
	This architecture uses a PBM augmentation module to predict unsampled multiuser PBMs and uses a rate mapping module to estimate the sum-rate corresponding to these predicted PBMs, which guides selecting better probing beams in the probing beam optimization module. 
	For ease of presentation, we form $BU$ vectors, i.e., $\boldsymbol{r}_{b,u}, b\in \mathbb{B}, u\in\mathbb{U}$, into a long vector $\boldsymbol{r}\in\mathbb{R}^{UN}$ with $N\triangleq \sum_{b=1}^{B}N_b$.
	$\mathbb{L}^\mathrm{sampled} \subseteq \mathbb{L}^\mathrm{total}$ denotes the set of indices of sampled probing beam combinations.
	The procedure of this framework is explained in the following.

	\textbf{The training phase}:
	\textit{Step 1:}
	The AP side collects the PBMs $\boldsymbol{r}^{(n)}|\tilde{\mathbb{I}}^{(l)}$ and the sum-rate $R_\mathrm{sum}^{(n)}|\tilde{\mathbb{I}}^{(l)}$, $n=1,...,\tilde{N}_l$, $l\in \mathbb{L}^\mathrm{sampled} $ for $\tilde{N}_l$ sampled channel realizations and $|\mathbb{L}^\mathrm{sampled}|$ sampled probing beam combinations.
	\textit{Step 2:} The PBM augmentation module utilizes the sampled PBMs
	$\boldsymbol{r}^{(n)}|\tilde{\mathbb{I}}^{(l)}$, $n=1,...,\tilde{N}_l$, $l\in \mathbb{L}^\mathrm{sampled}$ to recover the distribution of PBM $\boldsymbol{r}$  given the probing beam combination $\tilde{\mathbb{I}}$.
	\textit{Step 3:}
	The rate mapping module utilizes the sampled PBMs
	$\boldsymbol{r}^{(n)}|\tilde{\mathbb{I}}^{(l)}$ and sum-rates $R_\mathrm{sum}^{(n)}|\tilde{\mathbb{I}}^{(l)}$, $n=1,...,\tilde{N}_l, l\in \mathbb{L}^\mathrm{sampled}$
	to train the mapping $\{\boldsymbol{r},\tilde{\mathbb{I}}\} \rightarrow R_\mathrm{sum}$.
	
	\textbf{The execution phase}:
	\textit{Step 1:}
	The recovered PBM distribution during the training phase is used to sample a certain amount of augmented PBMs $\bar{\boldsymbol{r}}^{(n)}|\tilde{\mathbb{I}}^{(l)}$, $n=1,...,\tilde{N}^{\mathrm{aug}}_l$, $l\in \mathbb{L}^\mathrm{total}$.
	\textit{Step 2:}
	The augmented PBM $\bar{\boldsymbol{r}}^{(n)}|\tilde{\mathbb{I}}^{(l)}$, $n=1,...,\tilde{N}^{\mathrm{aug}}_l$ and the probing beam combination $\tilde{\mathbb{I}}^{(l)}$, $l\in \mathbb{L}^\mathrm{total}$ are input to the rate mapping module to obtain the sum-rate
	$\bar{R}_{\mathrm{sum}}^{(n)}|\tilde{\mathbb{I}}^{(l)}$, $n=1,...,\tilde{N}^{\mathrm{aug}}_l$, $l\in \mathbb{L}^\mathrm{total}$.
	\textit{Step 3:}
	Based on the original sampled data and the augmented data, the optimization module finds the optimal probing beam combination according to
	\setlength\abovedisplayskip{1pt}\setlength\belowdisplayskip{1pt}
	\begin{eqnarray}	
		\label{beam_chosen}
		\tilde{\mathbb{I}}^{(\star)}=\hspace{-0.3cm}\max\limits_{\tilde{\mathbb{I}}^{(l)}, \forall l\in \mathbb{L}^\mathrm{total}} \frac{\sum_{n=1}^{\tilde{N}_l}R_{\mathrm{sum}}^{(n)}\vert {\tilde{\mathbb{I}}}^{(l)} + \sum_{n=1}^{\tilde{N}_{l}^{\mathrm{aug}}}\bar{R}^{(n)}_{\mathrm{sum}}\vert \tilde{\mathbb{I}}^{(l)}}{\tilde{N}_l+\tilde{N}_l^{\mathrm{aug}}},
	\end{eqnarray}
	where $\tilde{N}_l=0$ for unsampled probing beam combination $l\in \mathbb{L}^\mathrm{total} - \mathbb{L}^\mathrm{sampled}$.

	\vspace{-0.4cm}
	\subsection{The Module Design}\label{structureofhe}
	\vspace{-0.1cm}
	Each module in Fig. \ref{ssb opt} is described in detail in the following.
	
	\subsubsection{PBM Augmentation Module}\label{dataaugmentation}
	
\begin{figure}[htbp]
		\vspace{-0.3cm}
		\centering
		\setlength{\abovecaptionskip}{0.cm}
		\includegraphics[scale=0.3]{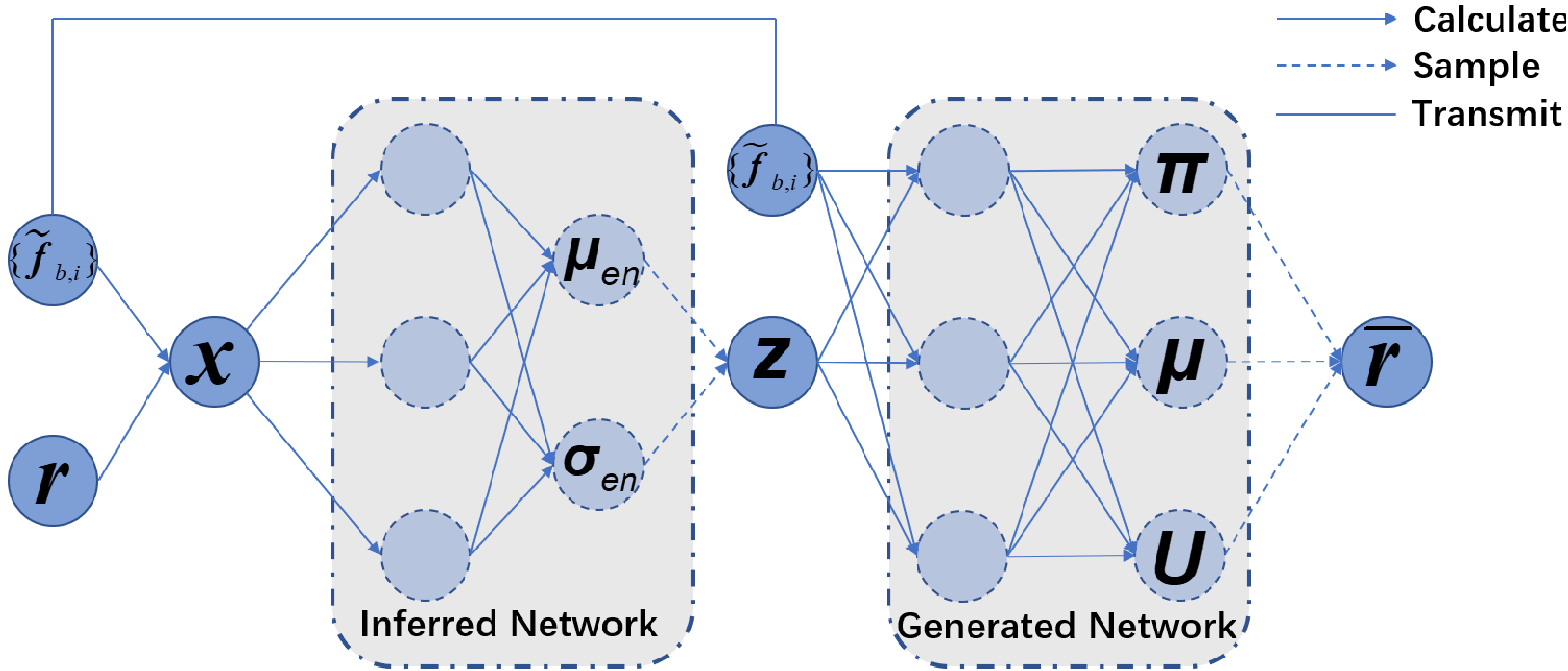}
		\caption{The structure of PBM Augmentation Module.}\label{cvae}
		\vspace{-0.5cm}
	\end{figure}

	We propose a CVAE-and-MDN-based PBM augmentation module to learn the PBM distribution under different probing beam combinations, which consists of an inferred network (encoder network) and a generated network (decoder network), as shown in Fig. \ref{cvae}.
	
	\textbf{In the training stage}, the input of inferred network composed of DNN is the codeword combination $\{\tilde{\boldsymbol{f}}^{(l)}_{b,i}\} \triangleq \{\tilde{\boldsymbol{f}}^{(l)}_{b,i}, \forall i\in\tilde{\mathbb{I}}^{(l)}_{b}, b\in \mathbb{B}\}$ and $\boldsymbol{r}^{(n)}|\tilde{\mathbb{I}}^{(l)}$, $n=1,...,\tilde{N}_l$, $l \in \mathbb{L}^{\mathrm{sampled}}$. 
	The output of inferred network are the mean $\boldsymbol{\mu}_\mathrm{en} \in \mathbb{R}^{N_{z}}$ and the standard deviation $\boldsymbol{\sigma}_\mathrm{en} \in \mathbb{R}^{N_{z}}$ of $N_z$ dimensional hidden Gaussian random vector $\boldsymbol z$, respectively. 
	If $\boldsymbol \varepsilon \sim {\mathcal{N}}(\mathbf 0, \mathbf I)$, then $\boldsymbol z$ can be sampled according to the output of inferred network as 
	$
	\boldsymbol z=\boldsymbol \mu_\mathrm{en}+ \boldsymbol \varepsilon \cdot \boldsymbol \sigma_\mathrm{en}.
	$
	Considering that the sampling of probing beam combination may be insufficient, i.e., $|\mathbb{L}^\mathrm{sampled}|<|\mathbb{L}^\mathrm{total}|$, it is better to use the codeword combination $\{\tilde{\boldsymbol{f}}_{b,i}\}$ rather than the combination index $\tilde{\mathbb{I}}_{b}, b\in\mathbb{B}$ as the condition variable, based on which we can better explore the out-of-distribution (OOD) generalization capability of the augmentation model.

	The generated network is composed of MDN, which makes it better represent the complex PBM distribution compared to the single multivariate Gaussian distribution, i.e.,
	\begin{equation} \label{MDNgaussian}
		p(\boldsymbol{r}|\tilde{\mathbb{I}}) = \sum_{g=1}^G \pi_g p_{g}(\boldsymbol{r}|\tilde{\mathbb{I}}),
	\end{equation}
	where $G$ is the number of distributions in the mixture.  $\pi_g$ satisfying $\sum_{g=1}^{G} \pi_{g}=1$ and $p_{g}(\boldsymbol{r}|\tilde{\mathbb{I}}), g=1,...,G$ are respectively the mixture portion and the probability density of the $g$-th multivariate Gaussian distribution, i.e.,
	$p_{g}(\boldsymbol{r}|\tilde{\mathbb{I}})\propto
	\exp\left(-\frac{1}{2}(\boldsymbol{r}-\boldsymbol{\mu}_{g})^\mathrm{T}\mathbf{\Sigma}_{g}^{-1}(\boldsymbol{r}-\boldsymbol{\mu}_{g})\right)\cdot|\mathbf{\Sigma}_{g}^{-1}|^{\frac{1}{2}}$.
	$\boldsymbol{\mu}_g\in\mathbb{R}^{UN}$ and $\boldsymbol{\Sigma}_{g}\in\mathbb{R}^{UN \times UN}, g=1,...,G$ are the mean vector and the covariance matrix of the $g$-th multivariate Gaussian distribution.
	\textit{The input of generated network} consists of the condition 
	$\{\tilde{\boldsymbol{f}}_{b,i}\}$ and the sampled hidden variable $\boldsymbol{z}|{\{\tilde{\boldsymbol{f}}_{b,i}\}}$. 
	Since the PBM is dependent on the user location given the probing beam combination, we consider the correlation among different PBM coefficients, i.e., $\boldsymbol{\Sigma}_g$ is a full covariance matrix rather than a diagonal one. 
	To calculate the probability in Eq. \eqref{MDNgaussian}, the key is to model $\boldsymbol{\Sigma}^{-1}_{g}$. However, it is difficult to guarantee the positive definite property of $\boldsymbol{\Sigma}^{-1}_{g}$ and the numerical computational stability via directly learning its coefficients. 
	Via drawing lessons from \cite{DBLP:journals/corr/abs-2003-05739}, we first conduct the Cholesky decomposition of $\boldsymbol{\Sigma}^{-1}_{g}$ and have $\boldsymbol{\Sigma}^{-1}_{g}=\mathrm{\boldsymbol{U}}_g^\mathrm{T}\mathrm{\boldsymbol{U}}_g$ with $\mathrm{\boldsymbol{U}}_g$ being an upper triangular matrix with positive diagonal elements.
	Therefore, \textit{the output of generated network} consists of $\boldsymbol{\pi}\!=\!\{\pi_g, g=1,...,G\}$, $\boldsymbol{\mu}\!=\!\{\boldsymbol{\mu}_g\in\mathbb{R}^{UN}, g=1,...,G\}$ and {$\boldsymbol{U}\!=\!\{\boldsymbol{U}_g\in\mathbb{R}^{UN \times UN}, g=1,...,G\}$. Then, we have the log-density of one mixture component as
		$\ln p_{g}(\boldsymbol{r}|\tilde{\mathbb{I}})
		\propto-\frac{1}{2}\left\|{\boldsymbol{U}}_g\left(\boldsymbol{r}-\boldsymbol{\mu}_g\right)\right\|_2^2+\sum_{j=1}^{UN} \ln {u}_{g,j}$
		where $\boldsymbol{u}_g=\operatorname{diag}({\boldsymbol{U}_g})$.
		
		Recall that the inferred network and the generated network in the CVAE-type network are both to maximize the evidence lower bound \cite{li2021influence}. For the proposed CVAE-and-MDN-based PBM augmentation module, the loss function is designed as
		$
		\text{Loss}_1= -\ln \sum_{g=1}^{G} \pi_{g}  p_{g}(\boldsymbol{r}|\tilde{\mathbb{I}}) + \mathrm{KL}\left({\mathcal{N}}(\mathbf{0}, \mathbf{I}), {\mathcal{N}}\left(\boldsymbol{\mu}_\text{en}, \boldsymbol{\sigma}_\text{en}\right)\right) 
		- \frac{1}{G}\sum_{g=1}^{G}\ln p_{g}(\boldsymbol{r}|\tilde{\mathbb{I}}),
		$
		where the first term is the log-likelihood function of the generated network output, the second term calculates the Kullback–Leibler (KL) divergence between the variational distribution ${\mathcal{N}}\left(\boldsymbol{\mu}_\text{en}, \boldsymbol{\sigma}_\text{en}\right)$ and the posterior distribution ${\mathcal{N}}(\mathbf{0}, \mathbf{I})$. Due to the high dimensionality of Gaussian distributions in our considered scenario (e.g., $144$ in our simulations), we introduce the third term to avoid the model degenerating to a single multivariate Gaussian distribution. 
			
			\textbf{In the execution stage}, only the generated network is called, whose inputs are $\{\tilde{\boldsymbol{f}}_{b,i}\}$ and the samples generated from the posterior distribution ${\mathcal{N}}(\mathbf{0}, \mathbf{I})$. 
			The augmentation module first selects the $g$-th multivariate Gaussian distribution according to the mixture portion $\boldsymbol{\pi}$, and then generates augmented PBM data $\bar{\boldsymbol{r}}\in\mathbb{R}^{UN}$ according to}
		$\bar{\boldsymbol{r}}\!=\!\boldsymbol{\mu}_g\!+\!\boldsymbol{U}^{-1}_{g}\boldsymbol{\varepsilon}$
		where $\boldsymbol{\varepsilon} \sim \mathcal{N}\left(\mathbf{0}, \mathbf{I}\right)$.
		
		\subsubsection{Rate Mapping Module}
		
		Typical advanced DNN networks, e.g., the DenseNet network, can be adopted for the rate mapping module. 
		During the training phase, the sampled PBMs 
		$\boldsymbol{r}^{(n)}|\tilde{\mathbb{I}}^{(l)}$, and corresponding sum-rates $R_\text{sum}^{(n)}|\tilde{\mathbb{I}}^{(l)}$,
		$n=1,...,\tilde{N}_l$ and $l\in \mathbb{L}^\text{sampled}$
		are inputted to the network $ f(\cdot)$ as data and labels, respectively.
		The minimum mean square error (MMSE) loss function is 
		\begin{equation} \label{MMSEloss}
			\begin{aligned}
				\text{Loss}_2 =  \left( \! {R_{\text{sum}}^{(n)}\vert \tilde{\mathbb{I}}^{(l)}}\! -\! f \left( {\boldsymbol{r}}\!^{(n)}\!|\tilde{\mathbb{I}}^{(l)}\!,\tilde{\mathbb{I}}^{(l)}\!;\!\boldsymbol{\theta}\!\right)\! \right)^2  
			\end{aligned}. 
		\end{equation}
		During the executing phase, the predicted sum-rates
		$\bar{R}_{\text{sum}}^{(n)}|\tilde{\mathbb{I}}^{(l)}$, $n=1,...,\tilde{N}^{\text{aug}}_l$, $l\in \mathbb{L}^\text{total}$
		are obtained by inputting the PBMs
		$\bar{\boldsymbol{r}}^{(n)}|\tilde{\mathbb{I}}^{(l)}$, $n=1,...,\tilde{N}^{\text{aug}}_l$, $l\in \mathbb{L}^\text{total}$
		and probing beam combinations
		$\tilde{\mathbb{I}}^{(l)}$, $l\in \mathbb{L}^\text{total}$
		into this module.

		\subsubsection{Beam Optimization Module}
		We propose a genetic algorithm \cite{mirjalili2019genetic}-based design. 
		First, the probing beam combinations
		$\tilde{\mathbb{I}}^{(l)}$, $l\in \mathbb{L}^\text{total}$
		are encoded to be individuals in the genetic space, and the maximum number of iterations $N_\text{it, max}$ is set. In each iteration, we set the maximum number of population evolutions $N_\text{ev, max}$ and randomly select some probing beam combinations to be encoded as the primary population. 
		Then, the following steps are repeated $N_\text{ev, max}$ times. The fitness of each individual in the contemporary population is calculated based on data augmentation and rate mapping modules, i.e,
		\begin{equation} \label{fitness}
			\begin{split}
				\text{Fitness}  =  -\frac{  \sum_{n=1}^{\tilde{N}_l} R_{\text{sum}}^{(n)} \vert {\tilde{\mathbb{I}}}^{(l)} +  
					\sum_{n=1}^{\tilde{N}_{l}^{\text{aug}}}  \bar{R}^{(n)}_{\text{sum}}\vert \tilde{\mathbb{I}}^{(l)}}{\tilde{N}_l+\tilde{N}_l^{\text{aug}}} 
			\end{split}. 
		\end{equation}
		Then the roulette method is utilized to select multiple individuals with the largest fitness 
		as the parent population for the next evolution. After that, the crossover operation and mutation operation, e.g.,  a single-point crossover operator and a single-point mutation operator, are used to act on the chosen high-quality population, respectively. At the end of the genetic algorithm, $\tilde{\mathbb{I}}^{(\star)}$ with the largest fitness is outputted.

		\vspace{-0.4cm}
		\section{Simulation and Discussion}
		\vspace{-0.20cm}
		The channel models are referred to Section \ref{Channel Model}. To ensure the reasonableness of parameter settings, we use the DeepMIMO channel dataset \cite{alkhateeb2019deepmimo} created by the commercial ray-tracing simulator. The 'O1' scenario is chosen and we select the area covered by rows R1000 to R2051 in the main street where three APs with indices 1, 3, 5 each with an $8 \times 8$ UPA serve 6 single-antenna users. 
		The system has a carrier frequency 28 GHz, an operating bandwidth of 100 MHz, and 1024 subcarriers. We sample one subcarrier to approximate a narrowband scenario. The transmit power of each AP in the sampled subcarrier is set to be 10 W.
		The beam training codebook is a UPA 2D-DFT matrix. 
		We set $N_b = 8$ for all APs and consider $\mathbb{L}^\mathrm{total}=8$ probing beam combinations. Specifically, we divide the beamspace into $8$ parts horizontally, i.e., $\tilde{\mathbb{I}}_b^{(l)}=\{(l-1)*8+0,...,(l-1)*8+7)\}, l\in \mathbb{L}^\mathrm{total}, b\in\mathbb{B}$.
		The area is divided into $6$ equal regions and the user locations are respectively randomly sampled within a square area with a side length of 4 meters around these regional centers. A total of $200$ sets of user locations are sampled. The sum-rate is calculated by Eq. \eqref{rate} with the hybrid BF in Section \ref{PBHBF}.

		We briefly illustrate the design of CVAE-and-MDN network in the PBM augmentation module, and that of the DNN network in the rate mapping module. 
		The total number of parameters in augmentation module input $\{\tilde{\boldsymbol{f}}^{(l)}_{b,i}\}$, $\boldsymbol{r}^{(n)}$, hidden variables $\boldsymbol{\mu}_\mathrm{en}, \boldsymbol{\sigma}_\mathrm{en}$ and output $\boldsymbol{\pi}$, $\boldsymbol{\mu}$, $\boldsymbol{U}$ are $1024$, $144$, $64$, $64$, $8$, $144$, $10440$, and that in rate-mapping module input $\{\tilde{\boldsymbol{f}}^{(l)}_{b,i}\}$, $\boldsymbol{r}^{(n)}$  and output $R_\text{sum}^{(n)}|\tilde{\mathbb{I}}^{(l)}$ are $1024$, $144$ and $1$. 
		The hidden layers of both modules consist of multiple linear layers, within each of which batch normalization, PReLU activation function, and dropout operation with dropout probability $0.3$ are employed. 
		The numbers of their total parameters are $11,651,562$ and $969,301$, respectively. The step-based learning rate schedule is adopted for training.
		And we set $N_\text{it, max}=3$ and $N_\text{ev, max}=5$ for beam optimization module.
		
		Since the KL divergence between the fitted distribution and the real distribution of PBM can not be directly calculated, we use the maximum mean discrepancy (MMD) to evaluate the performance of PBM augmentation module. Two distributions are identical when MMD equals zero \cite{gretton2012kernel}. Specifically, given $n$ examples from $P(x)$, $m$ samples from $Q(y)$, and the kernel function $k(\cdot)$, the MMD can be expressed as
		\begin{eqnarray} \label{mmd}
			\widehat{\operatorname{MMD}}(\!P, Q\!) &= \frac{\sum_{i=1}^n \sum_{j=1}^n k\left(X_i, X_j\right)}{n^2} + \frac{\sum_{i=1}^m \sum_{j=1}^m k\left(Y_i, Y_j\right)}{m^2} \nonumber \\
			&\hspace{2cm}-\frac{2\sum_{i=1}^n \sum_{j=1}^m k\left(X_i, Y_j\right)}{n m}.
		\end{eqnarray}
		The Gaussian kernel function is adopted and we consider the case $m=n$ for the calculation of MMD by Eq. \eqref{mmd} in the following. Besides the designed CVAE-and-MDN based PBM augmentation module, the modules based on CVAE and VAE-and-MDN are compared as baseline schemes. For VAE-and-MDN, $8$ networks are used for $8$ probing beam combinations.
		
\begin{figure*}[ht!]
			\vspace{-1.1cm}
			\centering
			\subfloat[]{\includegraphics[height=1.7in]{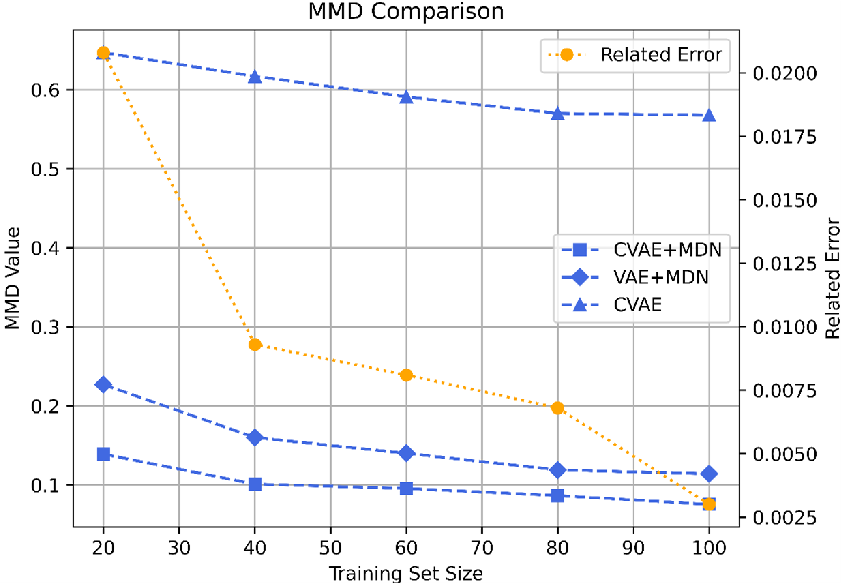}%
				\label{simu_1}}
			\hfil
			\subfloat[]{\includegraphics[height=1.7in]{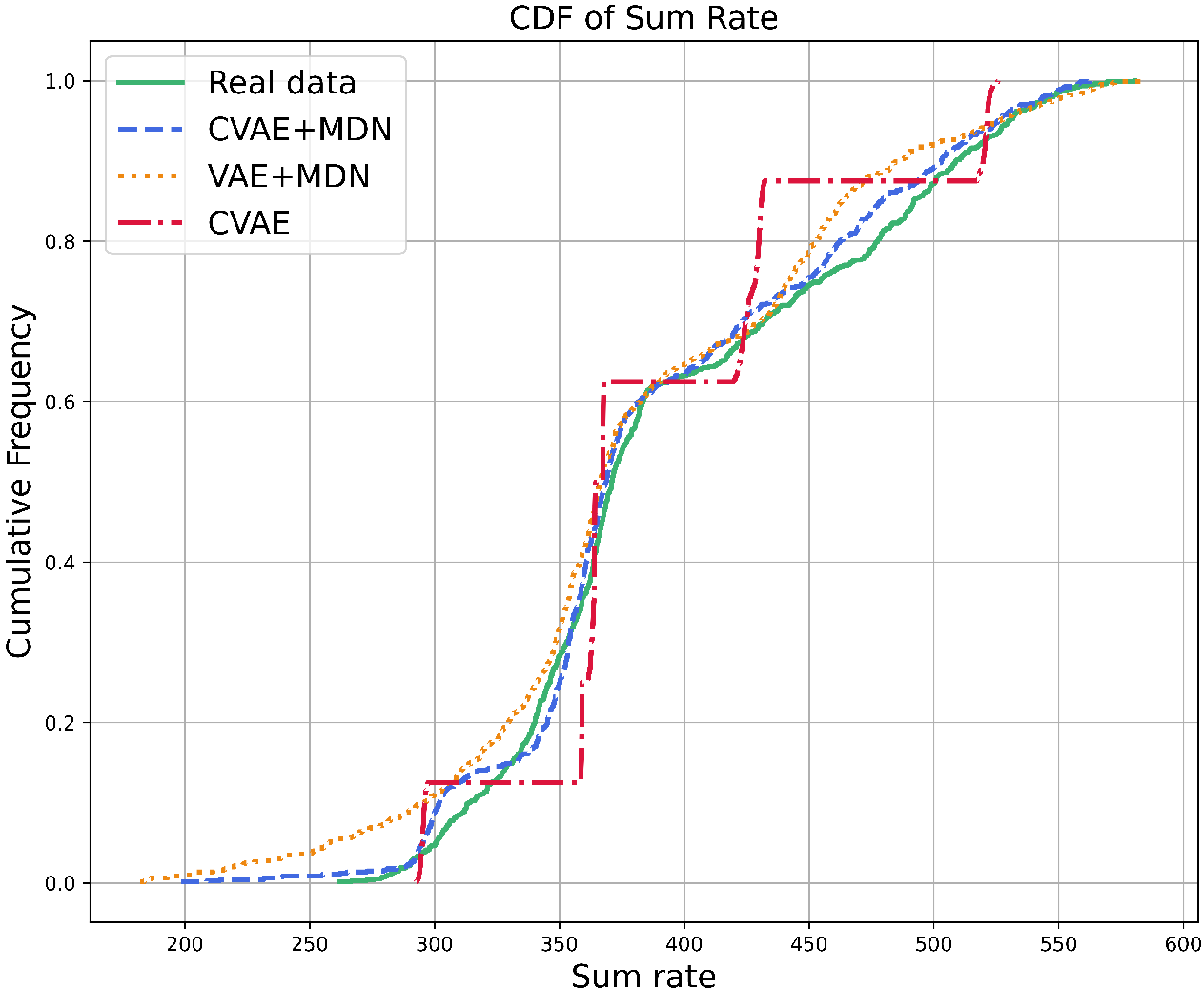}%
				\label{simu_3}}
			\hfil
			\subfloat[]{\includegraphics[height=1.7in]{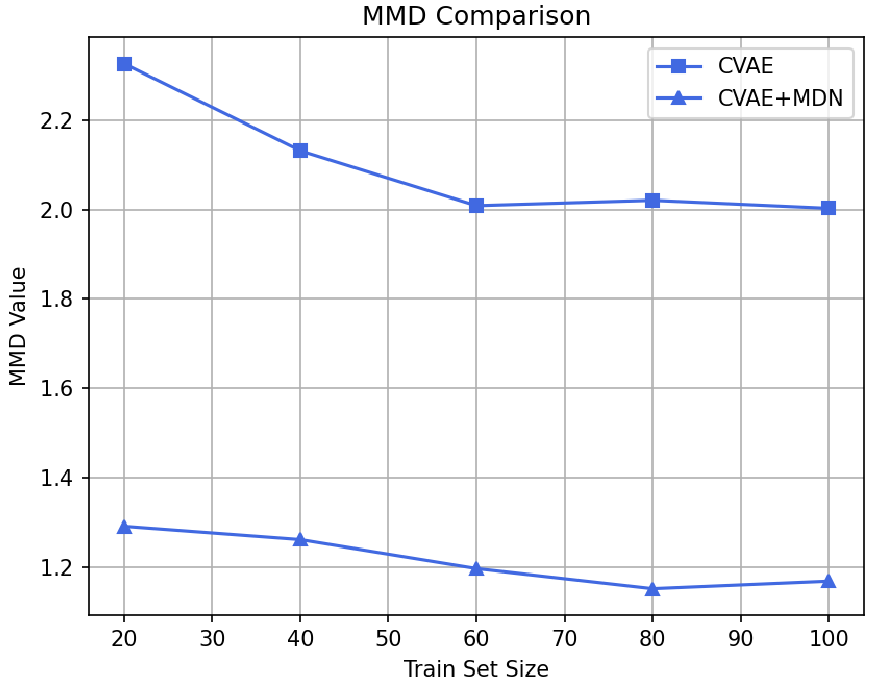}%
				\label{simu_2}}
			\vspace{-0.2cm}\caption{(a): The MMD of PBM augmentation with complete sampling in probing beam combination; (b): The CDF of predicted sum-rates for augmented PBMs; (c): The MMD of PBM augmentation with incomplete sampling in probing beam combination.}
			\label{fig_sim}
			\vspace{-0.6cm}
		\end{figure*}

		In Fig. \ref{simu_1} and \ref{simu_3}, the training data set involve samples under all $8$ probing beam combinations. The size of training set refers to the number of sampled PBMs for each probing beam combination. Fig. \ref{simu_1} shows that with test data size of $100$, the MMD decreases with more training samples for the proposed CVAE-and-MDN module and two benchmark modules. And the CVAE-and-MDN module outperforms the VAE-and-MDN module, both significantly outperforming the CVAE module. In addition, the designed rate mapping module provides accurate rate prediction for augmented PBMs, and the prediction error decreases quickly with more training data.
		\captionsetup[subfigure]{font=tiny}
		
		In Fig. \ref{simu_3}, the training and test set sizes are $40$ and $100$. 
		The CDFs of predicted sum rate powered by the proposed CVAE-and-MDN augmentation and the VAE-and-MDN benchmark are close to the real one, while the former has higher fitting accuracy. The sum rate CDF achieved by the CVAE benchmark is close to discrete distribution due to the mean-square error loss which could be detrimental for capturing variance. 
		Compared to CVAE, CVAE-and-MDN adopts multiple multivariate Gaussian distributions to have better learning capability, while the dimension of involved $\mathrm{\boldsymbol{U}}_g$ grows quadratically with the augmentation data dimension, which means the proposed CVAE-and-MDN module suffers greater training overhead and encounters higher risk of gradient vanishing when confronting with high-dimensional data augmentation tasks.
		Accordingly, the proposed probing beam optimization framework with the CVAE-and-MDN network consistently identify the optimal probing beam combination for $\tilde{N}_l = 20, 40, 60, 80, 100$ and $\tilde{N}_l^\text{aug} = 100$, while the one with the CVAE network outputs a sub-optimal combination for $\tilde{N}_l = 20$. 
		The average sum rate of probing beam-powered hybrid BF, under the optimal probing beam combination with a $75\%$ compression of the beamspace, achieves approximately $90\%$ of the sum rate attained by the SBF exhaustive search across the entire beamspace.
		
		In Fig. \ref{simu_2}, only $4$ probing beam combinations, i.e., $\tilde{\mathbb{I}}_b^{(l)}, l = 1,3,5,7$ are used in the data sampling. The remaining $4$ probing beam combinations only have $100$ test samples and no training samples. 
		It can be seen that both the CVAE network and the CVAE-and-MDN network have some capability to create PBMs under unsampled probing beam combinations. And the CVAE-and-MDN network achieves lower MMD value, showing its better applicability to scenarios where more possible probing beam combinations are considered.

		\vspace{-0.4cm}
		\section{Conclusion}
		\vspace{-0.2cm}
		In this letter, we proposed a probing beam optimization framework composed of three modules of PBM augmentation, sum-rate prediction and probing beam optimization. The proposed augmentation model, which organically integrates CVAE and MDN and generates multivariate Gaussian distributions with full covariance, is trained by the proposed Cholesky-decomposition based approach. Simulation results demonstrated that the proposed PBM augmentation model learns the PBM distribution more accurately than the traditional CVAE model, based on which the proposed probing beam optimization framework can select optimal probing beam combination effectively. In the future work, the probing beam optimization in finer temporal granularity, e.g, considering the impact of Rayleigh-type fast fading, should be studied.

		\vspace{-0.4cm}
		\bibliographystyle{IEEEtran}

\begin{thebibliography}{10}
			\vspace{-0.25cm}
			\providecommand{\url}[1]{#1}
			\csname url@samestyle\endcsname
			\providecommand{\newblock}{\relax}
			\providecommand{\bibinfo}[2]{#2}
			\providecommand{\BIBentrySTDinterwordspacing}{\spaceskip=0pt\relax}
			\providecommand{\BIBentryALTinterwordstretchfactor}{4}
			\providecommand{\BIBentryALTinterwordspacing}{\spaceskip=\fontdimen2\font plus
				\BIBentryALTinterwordstretchfactor\fontdimen3\font minus
				\fontdimen4\font\relax}
			\providecommand{\BIBforeignlanguage}[2]{{%
					\expandafter\ifx\csname l@#1\endcsname\relax
					\typeout{** WARNING: IEEEtran.bst: No hyphenation pattern has been}%
					\typeout{** loaded for the language `#1'. Using the pattern for}%
					\typeout{** the default language instead.}%
					\else
					\language=\csname l@#1\endcsname
					\fi
					#2}}
			\providecommand{\BIBdecl}{\relax}
			\BIBdecl
			
			\bibitem{wang2023full}
			D.~Wang, X.~You, Y.~Huang, \emph{et~al.}, ``Full-spectrum cell-free RAN for 6G systems: system
			design and experimental results,'' \emph{Sci. China Inf. Sci}, no.~3, p. 130305, 2023.
			
			\bibitem{10183795}
			X.~You, Y.~Huang, S.~Liu, \emph{et~al.}, ``Toward
			6G $\mbox{TK}\mu$ extreme connectivity: Architecture, key technologies and
			experiments,'' \emph{IEEE Wireless Commun.}, no.~3, pp.
			86--95, 2023.
			
			\bibitem{feng2021weighted}
			C.~Feng, W.~Shen, J.~An, and L.~Hanzo, ``Weighted sum rate maximization of the
			mmwave cell-free MIMO downlink relying on hybrid precoding,'' \emph{IEEE
				Trans. Wireless Commun.}, no.~4, pp. 2547-2560, 2021.
			
			\bibitem{wang2022joint}
			Z.~Wang, M.~Li, R.~Liu, and Q.~Liu, ``Joint user association and hybrid
			beamforming designs for cell-free mmwave MIMO communications,'' \emph{IEEE
				Trans. Commun.}, no. 11, pp. 7307-7321, 2022.
			
			\bibitem{yetis2021joint}
			C.~M. Yetis, E.~Bj{\"o}rnson, and P.~Giselsson, ``Joint analog beam selection
			and digital beamforming in millimeter wave cell-free massive MIMO systems,''
			\emph{IEEE Open J. Ind. Electron. Soc.}, pp. 1647--1662, 2021.
			
			\bibitem{hojatian2022decentralized}
			H.~Hojatian, J.~Nadal, J.-F. Frigon, and F.~Leduc-Primeau, ``Decentralized
			beamforming for cell-free massive MIMO with unsupervised learning,''
			\emph{IEEE Commun. Lett.}, no. 5, pp. 1042-1046, 2022.
			
			\bibitem{ma2020machine}
			W.~Ma, C.~Qi, and G.~Y. Li, ``Machine learning for beam alignment in millimeter
			wave massive MIMO,'' \emph{IEEE Wireless Commun. Lett.}, no.~6, pp.
			875--878, 2020.
			
			\bibitem{alkhateeb2018deep}
			A.~Alkhateeb, S.~Alex, P.~Varkey, \emph{et~al.}, ``Deep
			learning coordinated beamforming for highly-mobile millimeter wave systems,''
			\emph{IEEE Access}, pp. 37\,328--37\,348, 2018.
			
			\bibitem{heng2022learning}
			Y.~Heng, J.~Mo, and J.~G. Andrews, ``Learning site-specific probing beams for
			fast mmwave beam alignment,'' \emph{IEEE Trans. Wireless Commun.}, no.~8, pp.
			5785--5800, 2022.
			
			\bibitem{sun2018augmentation}
			W.~Sun, M.~Xue, H.~Yu, H.~Tang, and A.~Lin, ``Augmentation of fingerprints for
			indoor WIFI localization based on Gaussian process regression,'' \emph{IEEE
				Trans. Veh. Technol.}, no.~11, pp. 10\,896--10\,905, 2018.
			
			\bibitem{he2021intelligent}
			W.~He, C.~Zhang, Y.~Huang, and X.~You, ``Intelligent optimization of base
			station array orientations via scenario-specific modeling,'' \emph{IEEE
				Trans. Commun.}, no.~3, pp. 2117--2130, 2021.
			
			\bibitem{njima2021indoor}
			W.~Njima, M.~Chafii, A.~Chorti, R.~M. Shubair, and H.~V. Poor, ``Indoor
			localization using data augmentation via selective generative adversarial
			networks,'' \emph{IEEE Access}, pp. 98\,337--98\,347, 2021.
			
			\bibitem{chen2020progressive}
			L.~Chen, S.~Zhang, H.~Tan, and B.~Lv, ``Progressive RSS data augmenter with
			conditional adversarial networks,'' \emph{IEEE Access}, pp. 26\,975--26\,983,
			2020.
			
			\bibitem{ye2020deep}
			H.~Ye, L.~Liang, G.~Y. Li, and B.-H. Juang, ``Deep learning-based end-to-end
			wireless communication systems with conditional GANs as unknown channels,''
			\emph{IEEE Trans. Wireless Commun.}, no.~5, pp. 3133--3143, 2020.
			
			\bibitem{li2021influence}
			L.~Li, Z.~Zhang, and L.~Yang, ``Influence of autoencoder-based data
			augmentation on deep learning-based wireless communication,'' \emph{IEEE
				Wireless Commun. Letts}, no.~9, pp. 2090--2093, 2021.
			
			\bibitem{10255707}
			C.~Zhang, X.~Zhang, L.~Wei, \emph{et~al.}, ``Intelligent
			distributed beam selection for cell-free massive MIMO hybrid precoding,''
			\emph{IEEE Commun. Letts}, no.~11, pp. 2973--2977, 2023.
			
			\bibitem{orikumhi2020sinr}
			I.~Orikumhi, J.~Kang, H.~Jwa, J.-H. Na, and S.~Kim, ``Sinr maximization beam
			selection for mmwave beamspace mimo systems,'' \emph{IEEE Access}, vol.~8,
			pp. 185\,688--185\,697, 2020.
			
			\bibitem{mirjalili2019genetic}
			S.~Mirjalili and S.~Mirjalili, ``Genetic algorithm,'' \emph{Evol.
				Algorithms and Neural Netw.: Theory and Appl.}, pp. 43--55, 2019.
			
			\bibitem{alkhateeb2019deepmimo}
			A.~Alkhateeb, ``Deepmimo: A generic deep learning dataset for millimeter wave
			and massive MIMO applications,'' \emph{arXiv preprint arXiv:1902.06435},
			2019.
			
			\bibitem{hojatian2021unsupervised}
			H.~Hojatian, J.~Nadal, J.-F. Frigon, and F.~Leduc-Primeau, ``Unsupervised deep
			learning for massive MIMO hybrid beamforming,'' \emph{IEEE Trans.
				Wireless Commun.}, no.~11, pp. 7086--7099, 2021.
			
			\bibitem{DBLP:journals/corr/abs-2003-05739}
			J.~Kruse, ``Technical report: Training mixture density networks with full
			covariance matrices,'' \emph{arXiv preprint arXiv:2003.05739}, 2020.
			
			\bibitem{gretton2012kernel}
			A.~Gretton, K.~M. Borgwardt, M.~J. Rasch, \emph{et~al.}, ``A
			kernel two-sample test,'' \emph{J. Mach. Learn Res.},
			no.~1, pp. 723--773, 2012.
			
		\end{thebibliography}
		

	\end{document}